\documentclass[sigplan,screen,authorversion]{acmart}

\newif\ifcomments\commentsfalse   %% include author discussion
\newif\ifaftersubmission \aftersubmissionfalse %% Set true after submission
\newif\ifplentyofspace \plentyofspacetrue %% For stuff that we'd keep if we
\IfFileExists{texdirectives}{\input{texdirectives}}{}

\setcopyright{acmlicensed}
\acmPrice{15.00}
\acmDOI{10.1145/3293880.3294106}
\acmYear{2019}
\copyrightyear{2019}
\acmISBN{978-1-4503-6222-1/19/01}
\acmConference[CPP '19]{Proceedings of the 8th ACM SIGPLAN International Conference on Certified Programs and Proofs}{January 14--15, 2019}{Cascais, Portugal}
\acmBooktitle{Proceedings of the 8th ACM SIGPLAN International Conference on Certified Programs and Proofs (CPP '19), January 14--15, 2019, Cascais, Portugal}

\usepackage{booktabs}   %% For formal tables:
\usepackage{subcaption} %% For complex figures with subfigures/subcaptions
\usepackage{cleveref}

\usepackage{listings,lstlangcoq}

\usepackage{bussproofs}

\usepackage[export]{adjustbox}
\usepackage[normalem]{ulem}
\usepackage{balance}
\usepackage{breakurl}

\hypersetup{
  hyperindex=true,
  breaklinks=true
}

\lstdefinestyle{customc}{
  belowcaptionskip=1\baselineskip,
  breaklines=true,
  xleftmargin=\parindent,
  language=C,
  showstringspaces=false,
  basicstyle=\ttfamily,
  keywordstyle=\bfseries\color{green!40!black},
  commentstyle=\itshape\color{purple!40!black},
  identifierstyle=\color{blue!50!black},
  stringstyle=\color{orange},
}

\lstdefinestyle{customcoq}{
  mathescape=true,
  belowcaptionskip=1\baselineskip,
  breaklines=true,
  xleftmargin=\parindent,
  language=Coq,
  morekeywords={Variant, fun},
  emph={%
    SOCKAPI,ITree,data_at,data_at_
  },
  emphstyle={\bfseries\color{green!40!red!80}},
  showstringspaces=false,
  basicstyle=\small\ttfamily,
  keywordstyle=\bfseries\color{green!40!black},
  commentstyle=\itshape\color{purple!40!black},
  identifierstyle=\color{violet!80!black},
  stringstyle=\color{orange},
}

\ifcomments
\newcommand{\proposecut}[1]{\sout{#1}}
\newcommand{\ly}[1]{\textcolor{blue}{{LY:~#1}}}
\newcommand{\lx}[1]{\textcolor{olive}{{LX:~#1}}}
\newcommand{\bcp}[1]{\textcolor{violet}{{BCP:~#1}}}
\newcommand{\lys}[1]{\textcolor{green!50!black!100}{{LYS:~#1}}}
\newcommand{\nk}[1]{\textcolor{blue!50!green!100}{{NK:~#1}}}
\newcommand{\wm}[1]{\textcolor{orange}{{WM:~#1}}}
\newcommand{\wh}[1]{\textcolor{red!75!black!100}{{WH:~#1}}}
\newcommand{\sz}[1]{\textcolor{brown!100!black!100}{{SZ:~#1}}}
\newcommand{\lb}[1]{\textcolor{cyan!100!black!100}{{LB:~#1}}}
\else
\newcommand{\proposecut}[1]{}
\newcommand{\ly}[1]{}
\newcommand{\lx}[1]{}
\newcommand{\bcp}[1]{}
\newcommand{\lys}[1]{}
\newcommand{\nk}[1]{}
\newcommand{\wm}[1]{}
\newcommand{\wh}[1]{}
\newcommand{\sz}[1]{}
\newcommand{\lb}[1]{}
\fi

\newcommand{\op}{\mathtt{op}}
\newcommand{\inlinec}[1]{\lstinline[style=customc]{#1}}
\newcommand{\inlinecoq}[1]{\lstinline[style=customcoq,columns=flexible]{#1}}
\newcommand{\ITree}[1]{\inlinecoq{ITree(#1)}}

\newtheorem{theorem}{Theorem}

\begin{document}

%% Title information
% \title[Deeply Specifying a Networked Server]{Deeply Specifying a Networked
% Server}
\title{From C to Interaction Trees}
\subtitle{Specifying, Verifying, and Testing a Networked Server}
%\titlenote{with title note}             %% \titlenote is optional;
                                        %% can be repeated if necessary;
                                        %% contents suppressed with 'anonymous'
%\subtitle{Subtitle}                     %% \subtitle is optional
%\subtitlenote{with subtitle note}       %% \subtitlenote is optional;
                                        %% can be repeated if necessary;
                                        %% contents suppressed with 'anonymous'

%% Author information
%% Contents and number of authors suppressed with 'anonymous'.
%% Each author should be introduced by \author, followed by
%% \authornote (optional), \orcid (optional), \affiliation, and
%% \email.
%% An author may have multiple affiliations and/or emails; repeat the
%% appropriate command.
%% Many elements are not rendered, but should be provided for metadata
%% extraction tools.

\newcommand\pennauthor[1]{#1\textsuperscript\textdagger}

\author[Koh]{Nicolas Koh}
\affiliation{
  \institution{University of Pennsylvania}
  \city{Philadelphia}\state{PA}
  \country{USA}}
\author[Li]{Yao Li}
\affiliation{
  \institution{University of Pennsylvania}
  \city{Philadelphia}\state{PA}
  \country{USA}}
\author[Li]{Yishuai Li}
\affiliation{
  \institution{University of Pennsylvania}
  \city{Philadelphia}\state{PA}
  \country{USA}}
\author[Xia]{Li-yao Xia}
\affiliation{
  \institution{University of Pennsylvania}
  \city{Philadelphia}\state{PA}
  \country{USA}}
\author[Beringer]{Lennart Beringer}
\affiliation{
  \institution{Princeton University}
  \city{Princeton}\state{NJ}
  \country{USA}
}
\author[Honor\'{e}]{Wolf Honor\'{e}}
\affiliation{
  \institution{Yale University}
  \city{New Haven}\state{CT}
  \country{USA}
}
\author[Mansky]{William Mansky}
\affiliation{
  \institution{University of Illinois at Chicago}
  \city{Chicago}\state{IL}
  \country{USA}
}
\author[Pierce]{Benjamin C. Pierce}
\affiliation{
  \institution{University of Pennsylvania}
  \city{Philadelphia}\state{PA}
  \country{USA}}
\author[Zdancewic]{Steve Zdancewic}
\affiliation{
  \institution{University of Pennsylvania}
  \city{Philadelphia}\state{PA}
  \country{USA}}

\begin{abstract}
We present the first formal verification of a networked server implemented
in C.  {\em Interaction trees}, a general structure for representing
reactive computations, are used to tie together disparate verification and
testing tools (Coq, VST, and QuickChick) and to axiomatize the behavior of
the operating system on which the server runs (CertiKOS).  The main theorem
connects a specification of acceptable server behaviors, written in a
straightforward ``one client at a time'' style, with the
CompCert semantics of the C program.  The variability
introduced by low-level buffering of messages and interleaving of multiple
TCP connections is captured using {\em network refinement}, a variant of
observational refinement.
\end{abstract}

%% 2012 ACM Computing Classification System (CSS) concepts
%% Generate at 'http://dl.acm.org/ccs/ccs.cfm'.

%% End of generated code

%% Keywords
%% comma separated list

\maketitle
\sloppy

\section{Introduction}%\label{introduction-1-2-pages}

{\em The Science of Deep Specification}~\cite{deepspec} is an ambitious
experiment in specification, rigorous testing, and formal verification
``from internet RFCs down to transistors'' of
real-world systems such as web servers.
The principal challenges lie in integrating
disparate specification styles, legacy specifications, and testing and
verification tools to build and reason
about complex, multi-layered systems.

We report here on a first step toward realizing this vision: an in-depth case
study demonstrating how to specify, test, and verify a simple networked
server with the same fundamental interaction model as more sophisticated
ones---it communicates with multiple clients via ordered, reliable TCP
connections.  Our server is implemented in C and verified, using the
Verified Software Toolchain~\cite{vst}, against a formal ``implementation
model'' written in Coq~[\citeyear{coq}]; this is further verified (in Coq) against
a linear ``one client at a time'' specification of allowed behaviors.  The
main property we prove is that any trace that can be observed by a
collection of concurrent clients interacting with the server over the
network can be rearranged into a trace allowed by the linear
specification.  We also show how property-based random testing using Coq's
QuickChick plugin~\cite{quickchick} can be deployed in this setting, both
for detecting disagreements between the implementation and specification and for
validating the specification itself against legacy servers.
We compile the server code with the CompCert verified compiler~\cite{compcert}
and run it on CertiKOS~\cite{certikos}, a verified operating system with support
for TCP socket operations.

Our verified server provides a simple ``swap'' interface that allows clients to
send a new bytestring to the server and receive the currently stored one in
exchange.  It is simpler in many respects than a full-blown web server; in
particular, it follows a much simpler protocol (no authentication, encryption,
header parsing, \textit{etc.}), which means that it can be implemented with much
less code.

Moreover, the degree of vertical integration falls short of our ultimate
ambitions for the DeepSpec project, since we stop at the CertiKOS interface
(which we axiomatize) instead of going all the way down to transistors.
On the other hand, the C implementation of our server is realistic enough
that it offers a challenging test of how to integrate disparate Coq-based
methodologies and tools for verifying and testing systems software. In
particular, it uses a single-process, event-driven
architecture~\cite{flash}, hides latency by buffering interleaved TCP
communications from multiple clients, and is built on the POSIX socket API.

\paragraph{Contributions}

We describe our experiences integrating Coq, CompCert, VST, CertiKOS, and
QuickChick to build a verified swap server.  This is the first VST verification
of a program that interacts with the external environment. It is also, to the
best of our knowledge, the first verification of functional correctness of a networked server implemented in C.
Our technical contributions are as follows:

First, we identify {\em interaction trees} (ITrees)---a Coq adaptation of
structures known variously as ``freer''~\cite{freer},
``general''~\cite{mcbride-free}, or ``program''~\cite{freespec} monads---as a
suitable unifying structure for expressing and relating specifications at
different levels of abstraction (\Cref{sec:itrees}).

Second, we adapt standard notions of \emph{linearizability} and
\emph{observational refinement} from the literature on concurrent data
structures to give a simple specification methodology for networked servers
that is suitable both for rigorous property-based testing and for formal
verification.  We call this variant \emph{network refinement}
(\Cref{sec:network-refinement}).

Third, we demonstrate practical techniques for both \emph{verifying}
(\Cref{sec:verification}) and \emph{testing} (\Cref{sec:testing}) network
refinement between a low-level implementation model and a simple linear
specification.  We also demonstrate testing against the compiled C
implementation across a network interface.

Lastly, the ITrees embedded into both VST's separation logic and
CertiKOS's socket model allow us to make progress on
connecting the two developments. Though we leave completing the formal proofs
as future work, we identify the challenges and describe preliminary results
in~\Cref{sec:certikos}.

\Cref{overview} summarizes the whole development.
\Cref{sec:related-work,conclusion} discuss related and future work.

\section{Overview}\label{overview}

\begin{figure}[t]
  \includegraphics[width=.9\linewidth]{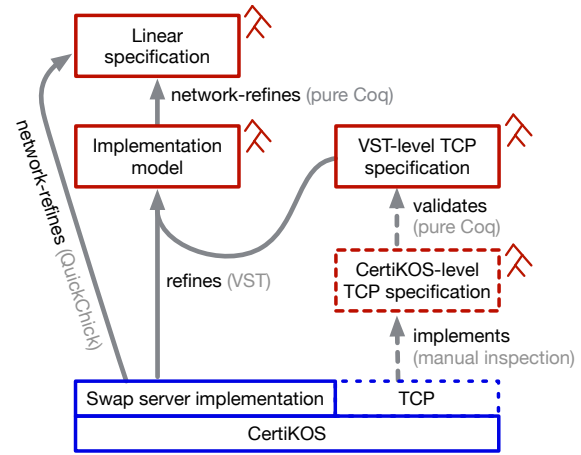}
\vspace*{-1.5ex}
  \caption{Overview. The blue parts of the figure
    represent components written in C, the red parts specifications in Coq.
    The swap server
    implementation runs on top of CertiKOS; it is proved to refine the
    implementation model with respect to a VST axiomatization of the socket
    interface; the axioms in VST, in turn, are validated by a lower-level
    axiomatization in the style of CertiKOS, which is manually compared to
    the (unverified) TCP implementation.  The implementation model
    ``network refines'' the linear specification.  The fact that
    the C implementation network refines the linear specification is
    independently validated by property-based random testing.  In all the
    Coq models and specifications, interaction trees model the observable
    behaviors of computations.  The dotted parts of the figure are either
    informal or incomplete.}
  \label{fig:overview}
\end{figure}

\begin{figure}[t]
  \centering
  \includegraphics[width=\linewidth,valign=t]{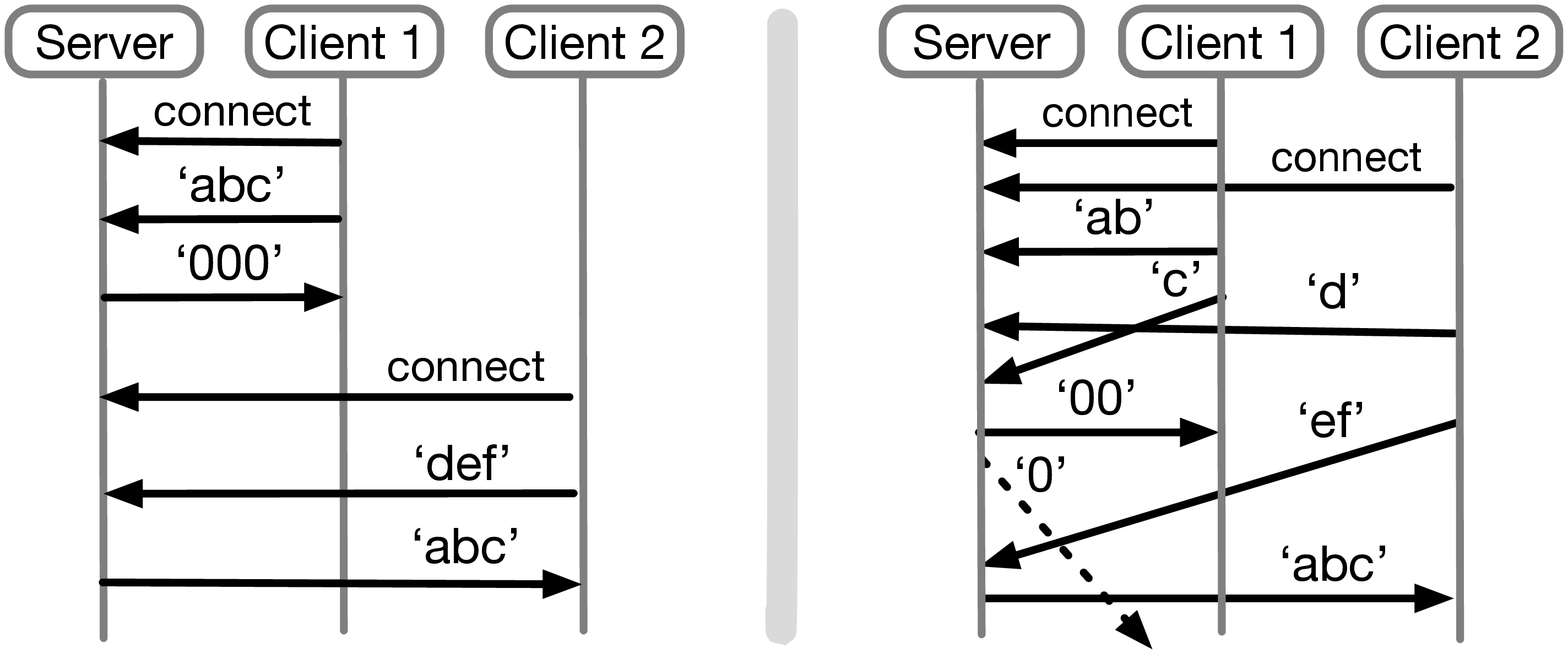}
\vspace*{-2ex}
  \caption{Swap server examples.  On the left is a simple run that
    directly illustrates the linear specification. Each client in turn
    establishes a connection, sends a three-byte message, and receives
    the message most recently stored on the server
    as a response. (`000' is the server's initial state.)
    On the right
    is another run illustrating internal buffering by the swap server
    and reordering by the network. The network socket may send one
    message in multiple chunks, messages from different clients may be
    received in any order, and messages may
    be delayed indefinitely (dotted arrow).  The ``explanation'' of
    the two runs in terms of the linear specification is the same.}
  \label{fig:swapserver2}
\end{figure}

Figure~\ref{fig:overview} shows the high-level architecture of the entire
case study.  This section surveys the major
components, starting with the high-level, user-facing specification (the linear specification shown at
the top of the figure) and working down to OS-level details.

\paragraph{Specifying the Swap Server}
\label{sec:top-level-theorem}
Informally, the intended behavior of the swap server is straightforward:
any number of clients can connect and send ``swap requests,'' each
containing a fixed-size message. The server acts as a one-element concurrent
buffer: it retains the most recent message that it has received and, upon
getting a swap request, updates its state with the new message and replies
to the sender with the old one.  The left-hand side of
Figure~\ref{fig:swapserver2} shows a simple example of correct behavior of a
swap server.

\begin{figure}[t]
\begin{lstlisting}[style=customcoq, basicstyle=\scriptsize\ttfamily]
CoFixpoint linear_spec' (conns : list connection_id)
           (last_msg : bytes) : itree specE unit :=
  or ( (* Accept a new connection. *)
       c <- obs_connect;;
       linear_spec' (c :: conns) last_msg )
     ( (* Exchange a pair of messages on a connection. *)
       c <- choose conns;;
       msg <- obs_msg_to_server c;;
       obs_msg_from_server c last_msg;;
       linear_spec' conns msg ).

Definition linear_spec := linear_spec' [] zeros.
\end{lstlisting}
\vspace*{-2ex}
\caption{Linear specification of the swap server.  In the
  \inlinecoq{linear_spec'} loop, the
  parameter \inlinecoq{conns} maintains the list of open connections, while
  \inlinecoq{last_msg} holds the message received from the last client
  (which will be sent back to the next client). \ifplentyofspace The server
  repeatedly chooses between accepting a new connection or doing a receive
  and then a send on some existing connection picked in the list
  \inlinecoq{conns}. \fi
  The linear specification is initialized with an empty set of connections
  and a message filled with zeros.}
\label{fig:linear-spec}
\end{figure}

\Cref{fig:linear-spec} shows the {linear} specification of the
server's behavior.  It says that the server can either accept a
connection with a new client (\inlinecoq{obs_connect}) or else receive
a message from a client over some established connection
(\inlinecoq{obs_msg_to_server} \inlinecoq{c}), send back the current
stored message (\inlinecoq{obs_msg_from_server} \inlinecoq{c}
\inlinecoq{last_msg}), and then start over with the last received
message as the current state.  The set of possible behaviors is
represented as an interaction tree (of type \inlinecoq{itree}
\inlinecoq{specE} \inlinecoq{unit}). We will discuss the types used
here in more details in \Cref{sec:itrees}.

Our main correctness theorem should relate the actual behavior of our server
(the CompCert semantics of the C code) to this linear description of its
desired behavior.  Informally:

\begin{theorem}
\label{thm:top-level-theorem}
Any sequence of interactions with the swap server that can be observed by
clients over the network could have been produced by the linear
specification.
\end{theorem}

This theorem constrains the implementation to act as a swap server: it prevents the
server from sending a message before it receives one, or while it has only
received a partial message; it prevents the server from sending an arbitrary value
in response to a request, or replying multiple times with the same value
that has only been received once; it also prevents the server from sending a response to
a client from which it has not received a request. However, the ``over the
network'' clause is a significant caveat: the server communicates with
clients via TCP, and even a correct implementation might thus exhibit a
number of undesirable behaviors from the clients' point of view. The network
might drop all packets after a certain point, causing the server to appear
to have stopped running, so the theorem allows the server to stop running at
any point. Similarly, the network might delay messages and might reorder
messages on different connections, so the theorem allows the server to
respond to an earlier request after responding to a later request.
\ifplentyofspace However, as long as the server performs any network
operations, those operations must be consistent with the protocol for a swap
server.  \fi The right-hand side of \Cref{fig:swapserver2} shows another run
of the system illustrating these possibilities; it should also be accepted
by the top-level theorem.

\begin{figure}[t]
\begin{lstlisting}[style=customcoq, basicstyle=\scriptsize\ttfamily]
  Theorem swap_server_correct :
    exists impl_model, ext_behavior C_prog impl_model /\
      network_refines linear_spec impl_model.
\end{lstlisting}
\vspace*{-2ex}
\caption{End-to-end swap server correctness theorem.}
\label{fig:servercorrect}
\end{figure}

Figure~\ref{fig:servercorrect} shows the formal specification
linking the \emph{linear specification} (\inlinecoq{linear_spec}), which
describes interactions with one client at a time, to the C program
(\inlinecoq{C_prog}). It is split in two parts articulated around
an \emph{implementation model} (\inlinecoq{impl_model}). It is
another interaction tree that describes the network-level behavior of the C
program more closely than the linear specification. Like the C program, the
implementation model interleaves requests from multiple clients and accounts
for the effects of the network.
A \emph{refinement} between the C program and the implementation model is
formalized by the VST property \inlinecoq{ext_behavior}. Then the
implementation model is connected to the specification by a different
\emph{network refinement} layer (\inlinecoq{network_refines}).

\paragraph{Network refinement}
The linear specification is short and easy to understand, but an
implementation that strictly followed it would be {\em obliged} to serve
clients sequentially, which is not what real servers (including ours) want
to do. Moreover, as shown on the right-hand side of
Figure~\ref{fig:swapserver2}, the network may delay and reorder messages, so
that, for example, the first two bytes of a message from client 1 might be
received after the first byte of a message from client 2. The server should
be able to account for this by buffering messages until they are complete.
The second part of our server specification loosens the linear specification
to account for the effects of communicating over a network; this also
permits realistic implementations that serve multiple clients concurrently.

Network refinement states that every possible behavior of the implementation
model is allowed by the linear specification, while accounting for message
reordering
and buffering that might be introduced by the network and/or server.
Section~\ref{sec:network-refinement} explains this process in more detail.

\paragraph{C Implementation}
\label{sec:c-implementation}

Our C implementation is a simple but reasonably performant server in a classical
single-process, event-driven style~\cite{flash}. The implementation
maintains a list of \inlinec{connection} structures, each representing a
state machine for one connection. Specifically, a connection structure
contains (1) a state, which may be \inlinec{RECVING}, \inlinec{SENDING}, or
\inlinec{DELETED}; (2) a buffer for storing bytes that have been received on
the connection; and (3) a buffer for storing bytes to send on the
connection.

The main body of the server is a non-terminating loop~(\Cref{fig:select-loop});
in each iteration, it uses the \inlinec{select} system
call\ifplentyofspace~\footnote{For simplicity, we choose \inlinec{select} over
  \inlinec{epoll}, a more efficient version found in Linux.}\fi{} to check for
pending connections to accept and for existing connections ready for
receiving/sending bytes from/to, and processes them. A new connection is handled
by initializing a new connection structure and adding it into the list, and an
existing connection is processed by updating the read/write buffers and
advancing the connection's state appropriately. This buffering strategy lets the
server interleave processing of multiple connections without having to wait for one client to send or receive a complete message.

\ifplentyofspace
{Our C code is compiled by
CompCert and should run on any operating system with POSIX
sockets.  We have tested it on CertiKOS, OSX, and Linux;
our long-term aim is deeper integration with CertiKOS's own
formal verification.}
\fi

\paragraph{Verifying the C code}

To prove that the C
implementation refines the implementation model (that is, that every possible network behavior of the C program is allowed by the implementation model), we use VST,
a tool for proving correctness of C programs using separation logic.  The
VST predicate \inlinecoq{ext_behavior C_prog impl_model} in
Figure~\ref{fig:servercorrect} relates the operational semantics of the C
program \inlinecoq{C_prog} to the interaction tree description given by
\inlinecoq{impl_model}.  Section~\ref{sec:impl-model} describes the
implementation model in more detail.

VST's model of program execution includes both conventional program state
(memory, local variables, \textit{etc.}) and \emph{external state}, an abstract
representation of the state of the environment in which the program is
  running. We connect the C program semantics to the implementation
model by adding a predicate \ITree{t} to VST's separation logic, asserting
that the environment expects the C program's network behavior
to match
the interaction tree \inlinecoq{t}.
\Cref{sec:verification}
describes this process.

\begin{figure}[t]
\begin{lstlisting}[style=customc, basicstyle=\scriptsize\ttfamily]
while(1 == 1) {
  ...
  int num_ready =
    select(maxsock + 1, &rs, &ws, &es, &timeout);
  if (num_ready <= 0) { continue; }
  int socket_ready = fd_isset_macro(server_socket, &rs);
  if (socket_ready) {
    /* Accept a new connection on the socket, create a
       connection structure for it, and link it into the
       head of the linked list of connections. */
    accept_connection(server_socket, &head);
  }
  /* For each connection in the list pointed to by head,
     read from or write to its buffer of data. */
  process_connections(head, &rs, &ws, last_msg_store);
}
\end{lstlisting}
\vspace*{-3.4ex}
  \caption{Main loop of swap server (in C).}
  \label{fig:select-loop}

\end{figure}

\paragraph{Assumptions and modeling gaps}
We have a complete
proof (using VST) that the C implementation compiled with CompCert
network-refines the linear specification---that is, a complete proof of the
claim in Figure~\ref{fig:servercorrect}. This proof is grounded in axiomatic
specifications of the OS-level system calls, and library functions like
\inlinec{memset} and the \inlinec{fdset} macros.
We rely on the soundness of the Coq proof assistant, plus the standard
axioms of functional and propositional extensionality and proof
irrelevance~\cite{coq-axioms}.

For this case study, our verification bottoms out at the interface between
the application program and the operating system; we rely on the correctness
of the OS's socket library and of the OS itself.  Since we are running on
CertiKOS, the OS has actually been proved correct, but its correctness
proofs and ours are not formally connected.  That is, our specification of
its socket API is axiomatized, but the axioms are partially validated by
connection to the corresponding CertiKOS specifications (specifically, a
VST specification of \inlinecoq{recv} has been partly connected to the
CertiKOS-level one; the other socket primitives remain to be connected). There
are several remaining challenges with connecting VST to CertiKOS, ranging
from the semantic---one critical technicality is connecting VST's
step-indexed view of memory with the flat memory model used by CertiKOS---to the technical---they use different versions of Coq.  See
Section~\ref{sec:certikos} for a fuller description of what we have
done to bridge these two formalizations.  Also, because CertiKOS currently
does not provide a verified TCP implementation, the best it can do is
mediate between the VST axioms and some, possibly lower-level,
axiomatization of the untrusted TCP stack. Filling these gaps is left to
future work.

\paragraph{Testing network refinement}
For our long-term goal of building verified systems software like web
servers, rigorous testing will be crucial, for two reasons.  First, even
small web servers are fairly complex programs, and they take significant effort
to verify; streamlining this effort by catching as many bugs as possible
before spending much time on verification makes good economic sense,
especially if the code can be automatically tested against the very same
specification that will later be used in the verification effort.  Second,
programs like web servers must often fit into an existing ecosystem---a
verified web server that interpreted the HTTP RFCs (\textit{e.g.},
\citet{rfc7540}) differently from Apache and Nginx would not be used.
Testing can be used to validate the formal specification against existing
implementations.

For the present case study, we use QuickChick~\cite{quickchick}, a Coq
plugin for property-based testing based on the popular QuickCheck
tool~\cite{quickcheck}.  We test both the compiled C code (by sending it
messages over a network interface) and the implementation model (by
exploring its behaviors within Coq) against the linear specification.

Supporting property-based testing requires \textit{executable}
specifications of the properties involved.  Happily, interaction trees,
which play a crucial role throughout our development, also work well with
Coq-style program extraction, and hence with testing.  Testing must also be
performed ``modulo network refinement'' in the same way as verification.
\Cref{sec:testing} describes this in more detail.

\section{Interaction Trees}\label{sec:itrees}

\begin{figure}[t]

  \begin{lstlisting}[style=customcoq, basicstyle=\scriptsize\ttfamily]
CoInductive itree (E : Type -> Type) (R : Type) :=
| Ret (r : R)
| Vis {X : Type} (e : E X) (k : X -> itree E R)
| Tau (t : itree E R).

Inductive event (E : Type -> Type) : Type :=
| Event : forall X, E X -> X -> event E.

Definition trace E := list (event E)

Inductive is_trace E R
  : itree E R -> trace E -> option R -> Prop := ...
  (* straightforward definition omitted *)
 \end{lstlisting}
\vspace*{-2.3ex}
  \caption{Interaction trees and their traces of events.}
  \label{fig:itrees}
\end{figure}

Components that interact with their environments appear at many levels in our
development (see \Cref{fig:overview}).  We use \textit{interaction trees}
(ITrees) as a general-purpose structure for specifying such components.
ITrees are a Coq adaptation of similar concepts known variously as
``freer,'' ``general,'' or ``program'' monads~\cite{freer, mcbride-free,
  freespec}.  We defer a deeper comparison until \Cref{sec:related-work}.

\paragraph{Constructing ITrees}\label{introductiondefinition}

Figure~\ref{fig:itrees} defines the type \inlinecoq{itree E R}.  The
definition is \textit{coinductive}, so that it can represent potentially
infinite sequences of interactions, as well as divergent behaviors.  The
parameter \inlinecoq{E} is a type of \textit{external interactions}---it
defines the interface by which a computation interacts with its environment.
\inlinecoq{R} is the \textit{result} of the computation: if the computation
halts, it returns a value of type \inlinecoq{R}.

There are three ways to construct an ITree. The \inlinecoq{Ret r} constructor
corresponds to the trivial computation that halts and yields the value
\inlinecoq{r}. The \inlinecoq{Tau t} constructor corresponds to a silent step of
computation, which does something internal that does not produce any visible
effect and then continues as \inlinecoq{t}.
Representing silent steps explicitly with \inlinecoq{Tau}
allows us, for example, to represent diverging computation
without violating Coq's guardedness condition~\cite{coinduction}:

  \begin{lstlisting}[style=customcoq, basicstyle=\scriptsize\ttfamily]
CoFixpoint spin {E R} : itree E R := Tau spin.
 \end{lstlisting}

The final, and most interesting, way to build an ITree is with the
\inlinecoq{Vis X e k} constructor.  Here, \inlinecoq{e : E X} is a ``visible''
external effect (including any outputs provided by the computation to its
environment) and \inlinecoq{X} is the type of data that the environment provides in
response to the event.  The constructor also specifies a continuation,
\inlinecoq{k}, which produces the rest of the computation given the
response from the environment.  \inlinecoq{Vis} creates branches
in the interaction tree because \inlinecoq{k} can behave differently for
distinct values of type \inlinecoq{X}.

Here is a small example that defines a type \inlinecoq{IO} of output or input
interactions, each of which works with natural numbers.  It is then
straightforward to define an ITree computation that loops forever, echoing
each input received to the output:

  \begin{lstlisting}[style=customcoq, basicstyle=\scriptsize\ttfamily]
Variant IO : Type -> Type :=
| Input  : IO nat
| Output : nat -> IO ().

CoInductive echo : itree IO () :=
  Vis Input (fun x => Vis (Output x) (fun _ => echo)).
 \end{lstlisting}

\paragraph{Working with ITrees}

Several properties of ITrees make them appealing as a structure for
representing interactive computations.  First, they are \textit{generic} in
the sense that, by varying the \inlinecoq{E} parameter, they can be
instantiated to work with different external interfaces.  Moreover, such
interfaces can be built compositionally: for example, we can combine a
computation with external effects in \inlinecoq{E1} with a different
computation with effects in \inlinecoq{E2}, yielding a computation with
effects in \inlinecoq{E1 +
  E2}, the disjoint union of \inlinecoq{E1} and \inlinecoq{E2}; there
is a natural inclusion of ITrees with interface \inlinecoq{E1} into ITrees
with interface \inlinecoq{E1 + E2}.
This approach is reminiscent of \textit{algebraic
  effects}~\cite{algebraic-effects}.  Our development exploits this
flexibility to easily combine generic functionality, such as a
nondeterministic choice effect (which provides the \inlinecoq{or} operator used
by the linear specification of Figure~\ref{fig:linear-spec}) with domain-specific
interactions such as the network send and receive events.  As with algebraic
effects, we can write a \textit{handler} or \textit{interpreter} for some
or all of the external interactions in an interface, for example to narrow the
effects \inlinecoq{E1 + E2} down to just those in \inlinecoq{E1}.  Typically,
such a handler will process the events of \inlinecoq{E2} and ``internalize''
them by replacing them with \inlinecoq{Tau} steps.

Second, the type \inlinecoq{itree E} is a
{monad}~\cite{MoggiMonads89,monad}, which makes it convenient to
structure effectful computations using the conventions and notations of
functional programming.  We wrap the \inlinecoq{Ret} constructor as
a \inlinecoq{ret} (return) function and use the sequencing notation
\inlinecoq{x <- e ;; k} for the monad's bind.  With a bit of wrapping and a
loop combinator \inlinecoq{forever}, we can rewrite the echo example with
less syntactic clutter:

\begin{lstlisting}[style=customcoq, basicstyle=\scriptsize\ttfamily]
  Definition echo : itree IO () :=
    forever (x <- input ;; output x)
\end{lstlisting}

Third, the ITree definition works well with Coq's extraction mechanism, allowing
us to represent computations as ITrees and run them for testing purposes.  Here
again, the ability to provide a separate interpretation of events is useful,
since its meaning can be defined outside of Coq.  In the echo example,
\inlinecoq{Output} events could be linked to a console output or to an OS's
network-send system call.  ITrees thus provide \textit{executable}
specifications.

One could, of course, simply consider such an extracted implementation to be the
final artifact (as in, for example, Verdi~\cite{verdi}).  However, we are
interested in a verified C implementation for two main reasons. First,
extracting Coq to OCaml generally involves a certain amount of
hackery---substituting native OCaml data structures for less efficient Coq ones,
interfacing with low-level operations such as I/O system calls,
\textit{etc.}---and this process is entirely unverified. Moreover, the extracted
code relies on OCaml's runtime and foreign-function interfaces, both of which
would have to be formalized to obtain the same strong guarantees that we hope to
achieve by connecting via C to CertiKOS.~\footnote{Compiling directly to native code using
CertiCoq~\cite{anand2017certicoq} would alleviate at least some of these
concerns.} Second, there is a potential performance gain from programming
directly in a low-level imperative language that may, in the long run, be
important for our eventual goal of verifying a high-performance web-server.

\begin{figure}[t]
  \begin{tabular}{rclr}
    \inlinecoq{r <- ei ;; k} & $\sqsubseteq$ &
    \inlinecoq{r <- or e1 e2 ;; k} & \inlinecoq{i} $\in \{1, 2\}$
    \\
    \inlinecoq{k x} & $\sqsubseteq$ &
    \inlinecoq{r <- choose l ;; k} & \inlinecoq{x} $\in$ \inlinecoq{l}
    \\
    \inlinecoq{r <- ret e ;; k} & $\equiv$ & \inlinecoq{k e}
    \\
    \inlinecoq{Tau k} & $\equiv$ & \inlinecoq{k}
    \\
    \multicolumn{4}{l}{      \inlinecoq{b <- (a <- e ;; f a) ;; g b}  $\equiv$
      \inlinecoq{a <- e ;; b <-  f a ;; g b}}
      \\
    \end{tabular}
\vspace*{-1.5ex}
    \caption{Trace refinement and equivalence for ITrees.}
    \label{fig:itree-props}
\end{figure}

\paragraph{Equivalence and Refinement}

Intuitively, ITrees that encode the same computation should be considered
equivalent. In particular, we want to equate ITrees that agree on their
terminal behavior (they return the same value) and on \inlinecoq{Vis}
events; they may differ by inserting or removing any finite number of
\inlinecoq{Tau} constructors. This ``{equivalence up to \inlinecoq{Tau}}''
is a form of {weak bisimulation}. We write \inlinecoq{t} $\equiv$
\inlinecoq{u} when \inlinecoq{t} and \inlinecoq{u} are equivalent up to
\inlinecoq{Tau}.  The monad laws for ITrees also hold modulo this notion of
equivalence. (Some of the laws used in our development are shown in
Figure~\ref{fig:itree-props}.)

ITrees that contain nondeterministic effects or that receive inputs from the
environment denote a {set} of possible \textit{traces}---(finite prefixes
of) execution sequences that record each visible event together with the
environment's response.  The definitions of \inlinecoq{trace} and the
predicate \inlinecoq{is_trace}, which asserts that a trace belongs to an
ITree, are shown in Figure~\ref{fig:itrees}.  Subset inclusion of behaviors
gives rise to a natural notion of ITree \textit{refinement}, written
\inlinecoq{t} $\sqsubseteq$ \inlinecoq{u}, which says that the traces of
\inlinecoq{t} are a subset of those allowed by \inlinecoq{u}.  We use this
refinement relation to allow an implementation to exhibit fewer behaviors
than those permitted by its specification.  Note that \inlinecoq{t} $\equiv$
\inlinecoq{u} implies \inlinecoq{t} $\sqsubseteq$ \inlinecoq{u}.

\paragraph*{ITrees as specifications: the linear specification }\label{specifying-system-programs}

Interaction trees provide a convenient yet rigorous way of formalizing
specifications.  We have already seen them in the linear specification of
the swap server in Figure~\ref{fig:linear-spec}.  The %
\inlinecoq{itree} \inlinecoq{specE} type there is an instance of
\inlinecoq{itree} whose visible events include nondeterministic choice as
well as observations of swap request and response messages, which are events
that include message content and connection ID information.  The
specification itself looks like a standard functional program that uses an
effect monad to capture network interactions.

\begin{figure}[t]
\begin{lstlisting}[style=customcoq, basicstyle=\scriptsize\ttfamily]
Definition select_loop_body
  (server_addr : endpoint_id)
  (buffer_size : Z)
  (server_st : list connection * string)
  : itree implE (bool * (list connection * string)) :=
  let '(conns, last_full_msg) := server_st in
  or
    (r <- accept_connection server_addr ;;
     match r with
     | Some c => ret (true, (c::conns, last_full_msg))
     | None   => ret (true, (conns, last_full_msg)) end)
    (let waiting_to_recv :=
         filter (has_conn_state RECVING) conns in
     let waiting_to_send :=
         filter (has_conn_state SENDING) conns in
     c <- choose (waiting_to_recv++waiting_to_send);;
     new_st <- process_conn buffer_size c last_full_msg;;
     let '(c', last_full_msg') := new_st in
     let conns' :=
         replace_when
           (fun x => if (has_conn_state RECVING x
                      || has_conn_state SENDING x)%bool
              then (conn_id x = conn_id c' ?)
              else false) c' conns in
     ret (true, (conns', last_full_msg'))).
\end{lstlisting}
\vspace*{-2ex}
\caption{Loop body of the implementation model.}
\label{fig:select-loop-model}
\end{figure}

\paragraph{ITrees as specifications: the implementation model}
\label{sec:impl-model}
We use the same \inlinecoq{itree} datatype, this time instantiated with an event
type \inlinecoq{implE} which contains nondeterministic choice and a networking
interface (\textit{e.g.}, \inlinecoq{accept}, \inlinecoq{send}, \inlinecoq{recv}), to define
the implementation model, which is a lower-level (but still purely functional)
specification of the swap server that more closely resembles the C code.
Figure~\ref{fig:select-loop-model} shows the body of the main loop from the
implementation model.

In contrast to the linear specification, the implementation model maintains
a list of connection structures instead of bare connection identifiers. Each
structure records the state for some connection. The state indicates whether
the server should be \inlinecoq{SENDING} or \inlinecoq{RECVING} on the
connection (or whether the connection is closed).  The state also records
the contents of send and receive buffers.  In each iteration of the loop, the
server either accepts a new connection or services a connection that is in
the \inlinecoq{SENDING} or \inlinecoq{RECVING} state. Servicing a
connection in the \inlinecoq{SENDING} state means sending some prefix of the
bytes in the send buffer; servicing a connection in the \inlinecoq{RECVING}
state means receiving some bytes on the connection.

Note that the control flow of this model differs from both the linear specification and
the C implementation.  The linear specification bundles together request--response pairs
and totally abstracts away from the details of buffering and interleaving communications
among multiple clients.  The relationship between the implementation model and
the linear specification is given by \textit{network refinement}, as we explain in the
next section.  For the C implementation, a single iteration of the main server
loop in Figure~\ref{fig:select-loop} corresponds to multiple
iterations of the select loop body of the model.  Nevertheless, we can
prove that the C behavior is a refinement of the implementation model, as we
describe in Section~\ref{sec:c-refines-model}.

\section{Network Refinement}\label{sec:network-refinement}

We show a ``network refinement'' relation between the implementation model
and the linear specification. At a high level, this property is a form of
{\em observational refinement}~\cite{data-refinement}: the behaviors of the
implementation that can be observed from across the network are included in
those of the specification. Intuitively, this property is also an analog, in
the network setting, of \emph{linearizability} for concurrent data
structures; we compare them in detail in Section~\ref{sec:related-work}.

\paragraph{The network}

We model a simple subset of the TCP socket interface, where connections
carry bytestreams (the bytes sent on an individual connection are ordered);
they are bidirectional (both ends can send bytes) and reliable (what is
received is a prefix of what was sent). This network model is represented by a
nondeterministic state machine where each connection carries a pair of
buffers of ``in flight'' bytes, with labeled transitions for a
client to open a connection, a server to accept it, and either party to send
and receive bytes (Figures~\ref{fig:networkevent} and~\ref{fig:networktransition}).
\begin{figure}[t]
\begin{lstlisting}[style=customcoq, basicstyle=\scriptsize\ttfamily]
Inductive network_event : Type :=
| NewConnection (c : connection_id)
| ToServer      (c : connection_id) (b : byte)
| FromServer    (c : connection_id) (b : byte).

Definition network_trace : Type := list network_event.
\end{lstlisting}
\vspace*{-2ex}
  \caption{Types for events and traces observed over the network.
  \inlinecoq{network_event} maps to \inlinecoq{event} values to form
  traces for both the specification and the implementation model.}
  \label{fig:networkevent}
\end{figure}
\begin{figure}[t]
\begin{lstlisting}[style=customcoq, basicstyle=\scriptsize\ttfamily]
Definition server_transition (ev : network_event)
    (ns ns' : network_state) : Prop :=
  match ev with
  | FromServer c b => let cs := Map.lookup ns c in
    match connection_status cs with
    | ACCEPTED  => let cs' := update_out
                    (connection_outbytes cs ++ [b]) cs
                  in ns' = Map.update c cs' ns
    | PENDING | CLOSED => False end
  | ... (* Other two cases *) end.

Definition client_transition : network_event ->
  network_state -> network_state -> Prop := ...
\end{lstlisting}
\vspace*{-2ex}
  \caption{Network transitions labeled by \inlinecoq{network_event},
  showing only the case where the server sends a byte.}
  \label{fig:networktransition}
\end{figure}
For example, there is a transition from network state \inlinecoq{ns} to
state \inlinecoq{ns'},
labeled \inlinecoq{FromServer c b}, if the connection \inlinecoq{c}
was previously accepted by the server (its status in \inlinecoq{ns} is
\inlinecoq{ACCEPTED}) and the state \inlinecoq{ns'} is obtained from
\inlinecoq{ns} by adding byte \inlinecoq{b} to the outgoing bytes on
connection \inlinecoq{c}.

We define a relation \inlinecoq{network_reordered_ ns ts tc : Prop} between
server- and client-side traces of network events \inlinecoq{ts} and
\inlinecoq{tc}, which holds
if they can be produced by an execution of the network starting from state
\inlinecoq{ns}. For the initial state with all connections closed, we
define %
\inlinecoq{network_reordered} \inlinecoq{ts} \inlinecoq{tc} %
\inlinecoq{= network_reordered_ initial_ns ts tc}.
The trace \inlinecoq{tc} is a ``disordering'' of
\inlinecoq{ts}---\textit{i.e.,} \inlinecoq{tc} is one
possible trace a client may observe if the server generated the trace
\inlinecoq{ts}.
Conversely, \inlinecoq{ts} is a ``reordering'' of \inlinecoq{tc}.

\paragraph{Network behavior of ITrees}
As mentioned in \Cref{sec:itrees}, ITrees such as the implementation model (of
type \inlinecoq{itree} \inlinecoq{implE}) and the linear specification
(\inlinecoq{itree specE}) define sets of event traces.
From across the network, those events can appear \emph{disordered} to the client,
so the \emph{network behavior} of an ITree is the set of possible
disorderings of its traces (defined using \inlinecoq{network_reorder}).
Finally, the ITree \inlinecoq{impl_model} \emph{network refines} the
\inlinecoq{linear_spec} when the former's network behavior is included in the
latter's; see \Cref{fig:netrefinesDef}.

\begin{figure}[t]
\begin{lstlisting}[style=customcoq, basicstyle=\scriptsize\ttfamily]
Definition impl_behavior (impl : itree implE unit) : network_trace -> Prop :=
  fun tr => exists tr_impl, is_impl_trace impl tr_impl /\ network_reordered tr_impl tr.

Definition spec_behavior (spec : itree specE unit) : network_trace -> Prop :=
  fun tr => exists tr_spec, is_spec_trace spec tr_spec /\ network_reordered tr_spec tr.

Definition network_refines impl spec : Prop :=
  forall tr, impl_behavior impl tr -> spec_behavior spec tr.
\end{lstlisting}
\vspace*{-2ex}
  \caption{Definition of network refinement in Coq.
  The functions \inlinecoq{is_impl_trace} and \inlinecoq{is_spec_trace}
  are thin wrappers around \inlinecoq{is_trace} that convert between
  traces of different (but isomorphic) event types.}
  \label{fig:netrefinesDef}
\end{figure}

\paragraph{Proving network refinement}

In order to prove that our implementation model network refines the linear
specification, we establish logical proof rules for a generalization
of \inlinecoq{network_refines}, named \inlinecoq{nrefines_}
(Figure~\ref{fig:nrefinesDef}).
The \inlinecoq{nrefines\_}% this comment fixes a typesetting bug
relation is step-indexed (\inlinecoq{z : nat}) to handle the server's
nonterminating loop; it relates a subtree of the implementation model
\inlinecoq{impl} to a record \inlinecoq{s} of the current state of the network
(\inlinecoq{get\_ns s: network\_state}) and a subtree of the specification ITree
(\inlinecoq{get\_spec s : itree specE unit}).

\begin{figure}[t]
\begin{lstlisting}[style=customcoq, basicstyle=\scriptsize\ttfamily]
Record state := { get_ns : network_state;
                  get_spec : itree specE unit; ... }.

Definition nrefines_ (z : nat) (s : state)
                     (impl : itree implE unit) : Prop :=
  forall tr, is_impl_trace_ z s impl tr ->
    exists dstr : network_trace,
      network_reordered_ (get_ns s) dstr tr /\
      is_spec_trace (get_spec s) dstr.
\end{lstlisting}
\vspace*{-2ex}
  \caption{Refinement relation generalized for reasoning.}
  \label{fig:nrefinesDef}
\end{figure}

Two example proof rules are shown in \Cref{fig:nrefinesrules}.
When the server performs a network operation, for example when it receives a
byte on a connection \inlinecoq{c}, we use a lemma such as
\inlinecoq{nrefines_recv_byte_}: we must prove that
the connection \inlinecoq{c} is open, and we then prove the
\inlinecoq{nrefines_} relation on the continuation \inlinecoq{k b},
with an updated network state in \inlinecoq{s'}.

At any point in the proof, we can also generate a part of the reordered trace
from the linear specification ITree \inlinecoq{get_spec} \inlinecoq{s}, using
the \inlinecoq{nrefines_network_transition_} lemma. We actually use this rule
at exactly two ``linearization points'' in the implementation model: right after
the server accepts a new connection, and after it receives a complete message
from a client and swaps it with the last stored message.

% N.B:
% The typesetting here is very fragile. Please check the rendered result if you
% touch this paragraph especially spaces around inlinecoq
% (or implement a more robust solution!)
%
Using these rules, we prove the proposition \inlinecoq{forall z, nrefines_}% (This comment fixes a typesetting bug)
\inlinecoq{z s0 impl_model}, where \inlinecoq{s0} is defined so that
\inlinecoq{get_spec s0 =}% (This comment fixes a typesetting bug)
\inlinecoq{linear_spec} and \inlinecoq{get_ns s0} is the
initial network state, where all connections are closed;
we can show this implies the second clause of the correctness theorem
(\Cref{fig:servercorrect}).

\begin{figure}[t]
\begin{lstlisting}[style=customcoq, basicstyle=\scriptsize\ttfamily]
Lemma nrefines_recv_byte_ z s
  (c : connection_id) (k : byte -> itree implE unit)
  : In (get_status s c) [PENDING; ACCEPTED] ->
    (forall b s', s' = append_inbytes c [b] s ->
             nrefines_ z s' (k b)) ->
    nrefines_ z s (b <- recv_byte c;; k b).

Lemma nrefines_network_transition_ z s spec' ns' impl
      (dtr : network_trace)
  : (forall dtr', is_spec_trace spec' dtr' ->
             is_spec_trace (get_spec s)
             (dtr ++ dtr')) ->
    server_transitions dtr (get_ns s) ns' ->
    nrefines_ z (set_ns ns' (set_spec spec' s)) impl ->
    nrefines_ z s impl.
\end{lstlisting}
\vspace*{-2ex}
  \caption{Example proof rules for \inlinecoq{nrefines_}.}
  \label{fig:nrefinesrules}
\end{figure}

\section{Verification}\label{sec:verification}

\paragraph*{Embedding ITrees in VST}
\label{subsec:itrees-in-hoare}

VST is a framework for proving separation logic specifications of C programs, based on the C semantics of the CompCert compiler. Its separation logic comes with a proof
automation system, Floyd, that supplies tactics for symbolically executing a program while maintaining its pre- and postcondition~\cite{DBLP:journals/jar/CaoBGDA18}.
To support
reasoning about external behavior in general---and the swap server's
invocations of OS/network primitives in particular---we extend VST's logic with two
\emph{abstract predicates}~\cite{abstract-pred}; these are separation logic
predicates
that behave like resources but do not have a footprint in concrete
memory.  Instead they connect to VST's model of \emph{external state}, which in
this case represents the allowed network behavior of the program. To make this
possible, we made a small modification to the internals of VST to enable it to refer to the external state in assertions.

The first abstract predicate, \inlinecoq{ITree(t)}, injects an
interaction tree \inlinecoq{t} into a VST assertion (an \inlinecoq{mpred}):
\begin{lstlisting}[style=customcoq, basicstyle=\scriptsize\ttfamily]
  Definition ITree {R} (t : itree implE R) : mpred :=
    EX t' : itree implE R, !!(t $\sqsubseteq$ t') && has_ext t'.
\end{lstlisting}
\inlinecoq{ITree t} asserts that the observation traces of \inlinecoq{t}
(\textit{i.e.}, the traces that
may be produced by a program satisfying the assertion \inlinecoq{ITree t}) are
included in the traces that are permitted by the external environment (here,
the OS). The \inlinecoq{has_ext} predicate asserts that the external state (here representing the network behavior the OS expects from the program) is exactly \inlinecoq{t'}.
The notation \inlinecoq{!!p} lifts an ordinary
Coq predicate \inlinecoq{p} to a VST separation logic predicate, and
\inlinecoq{&&} and \inlinecoq{EX} are logical conjunction and existential quantification at the level of separation logic assertions.

While a detailed description of VST's support for external
state is beyond the scope of the present paper and will be reported
elsewhere, we give some key properties of
this embedding.  Internal code execution does not
depend on or alter external state, so every program step that is not a call to the socket API leaves the \inlinecoq{ITree} predicate unchanged.  The monad and equivalence laws from the
abstract theory of interaction trees are reflected as (provable) entailments
between \inlinecoq{ITree} predicates (recall the refinement relation of
Figure~\ref{fig:itree-props}):

\begin{prooftree}
    \AxiomC{\inlinecoq{t} $\sqsubseteq$ \inlinecoq{u}}
    \UnaryInfC{\inlinecoq{ITree u} $\vdash$ \inlinecoq{ITree t}}
\end{prooftree}

\noindent This rule is \textit{contravariant} because we can conform to
the ITree \inlinecoq{u} by producing some subset of its allowed behavior.

External calls to network and OS functions are equipped with specifications
that reflect the evolution of interaction trees, in resource-consuming
fashion: actions are ``peeled off'' from the ITree as execution proceeds, so
that the interaction tree in the postcondition of an external function
specification is a subtree of the tree in the
precondition.  The ITree found in the outermost precondition of a program is
thus a sound approximation of all the program's external interactions.

\paragraph{Hoare-logic specifications of system calls}

\begin{figure}[t]
\begin{lstlisting}[style=customcoq, basicstyle=\scriptsize\ttfamily]
  { SOCKAPI st * ITree t *
    data_at_ alloc_len buf_ptr *
    !! ((r <- recv client_conn (Z.to_nat alloc_len) ;; k r) $\sqsubseteq$ t) *
    !! (consistent_world st /\ lookup_socket st fd = ConnectedSocket client_conn) *
    !! (0 <= alloc_len <= SIZE_MAX) }
\end{lstlisting}
\vspace{-0.3cm}
\begin{lstlisting}[style=customc, basicstyle=\scriptsize\ttfamily]
  ret = recv(int fd, void* buf_ptr, unsigned int alloc_len, int flags)
\end{lstlisting}
\vspace{-0.3cm}
\begin{lstlisting}[style=customcoq, basicstyle=\scriptsize\ttfamily]
  { exists (result : unit + option string) st' ret contents,
    !! (0 $\leq$ ret $\leq$ alloc_len \/ ret = -1) *
    !! (ret > 0 -> (exists msg, result = inr (Some msg) /\ ...) /\ st' = st) *
    !! (ret = 0 -> result = inr None /\ ...) *
    !! (ret < 0 -> result = inl tt /\ ...) *
    !! (Zlength contents = alloc_len) *
    !! (consistent_world st') *
    SOCKAPI st' *
    ITree (match result with
    | inl tt => t
    | inr msg => k msg end) *
    data_at alloc_len contents buf_ptr}
\end{lstlisting}
\vspace*{-2ex}
\caption{VST axiom for the \inlinec{recv} system call.}
\label{fig:posix-hoare}
\end{figure}

This use of the \inlinecoq{ITree} predicate can be seen in the VST axiom for
the \inlinec{recv} system call in Figure~\ref{fig:posix-hoare}.  The
precondition of this rule requires that the ITree \inlinecoq{(r <- recv
  client_conn (...);; k r)}, which starts with a \inlinecoq{recv} event, be
among the allowed behaviors of \inlinecoq{t}, so a legal implementation of
this specification is allowed to perform a \inlinec{recv} call next.  The
postcondition either leaves the interaction tree \inlinecoq{t} untouched, in
the case that the call to \inlinec{recv} failed, or says that the
implementation may continue as \inlinecoq{k msg}, in the case that the call
to \inlinec{recv} successfully returned a message \inlinecoq{msg}.

Most of the remaining constraints relate the program variables and the variables in the interaction tree to the corresponding state in
memory. For example, the predicate \inlinecoq{data_at_ alloc_len buf_ptr}
says that \inlinecoq{buf_ptr} points to a buffer of length
\inlinecoq{alloc_len}. The constraint \inlinecoq{lookup_socket st fd =
  ConnectedSocket client_conn} says that the socket with identifier
\inlinecoq{fd} is in the {\sc connected} state according to the API and is
associated with the connection identifier \inlinecoq{client_conn} appearing
in the interaction tree.

This socket information is tracked by a second abstract predicate,
\inlinecoq{SOCKAPI(st)}, which asserts that the external socket API memory can be
abstracted as \inlinecoq{st}, mapping file descriptors to socket states {\sc
  closed}, {\sc opened}, {\sc bound}, {\sc listening}, or {\sc connected}. Bound
and listening states are associated with an endpoint identifier in the network
model, and connected states are associated with a connection identifier in the
network model.  The reason for modularly separating socket states from
interaction trees is that the latter describe truly external behavior while the
former concern the (private) contract between the server program and the OS.
Specifically, the functions for creating sockets, binding them to addresses, and
closing sockets (after shutdown) are not visible at the other end of the network
and are hence specified to only operate over \inlinecoq{SOCKAPI} abstract
predicates. In general, system calls like \inlinec{recv} that affect the
network state carry specifications of the form

\begin{lstlisting}[style=customcoq, basicstyle=\scriptsize\ttfamily]
   { SOCKAPI(st) * ITree (x <- $\op(a_1, \ldots)$; k x) * $\ldots$ }
  \end{lstlisting}
  \vspace{-0.3cm}
  \begin{lstlisting}[style=customc, basicstyle=\scriptsize\ttfamily]
   op(a1, ...)
  \end{lstlisting}
  \vspace{-0.3cm}
  \begin{lstlisting}[style=customcoq, basicstyle=\scriptsize\ttfamily]
   { EX st' t'. SOCKAPI(st') * ITree(t') * $\ldots$ $\land$
     ($\phi$(r) $\to$ t' = k r) $\land$ ($\lnot \phi$(r) $\to$ t' = t)}
  \end{lstlisting}
  where $\phi$ is a boolean predicate distinguishing ITree-advancing
  (successful) invocations from failed invocations (which leave
  the ITree unmodified), by inspection of the implicitly quantified
  return value \inlinecoq{r}.

\paragraph*{Verifying the C implementation}
\label{sec:c-refines-model}
Having defined the abstract predicates we need to describe the network
behavior of the server, we can now prove that the C implementation refines
the implementation model using VST's separation logic.
The goal is to prove that the implementation
model \inlinecoq{impl_model} is an \emph{envelope} around the possible
network behaviors of the C program, \textit{i.e.}, every execution of the C program
performs only the socket operations described in \inlinecoq{impl_model};
this is expressed by the predicate \inlinecoq{ext_behavior C_prog impl_model}.
This proof then composes with the network refinement proof between
\inlinecoq{impl_model} and the linear specification to give us the
main theorem in Figure~\ref{fig:servercorrect}.

We prove \inlinecoq{ext_behavior C_prog impl_model} by specifying and proving a
Hoare triple for each function in the C implementation. We begin with
axiomatized Hoare triples for the library functions, in particular those
from the POSIX socket API; these triples modify the \inlinecoq{SOCKAPI} state
and possibly consume operations from the \inlinecoq{ITree}, as described
above.

We then specify Hoare triples for functions in the program, including embedded
interaction trees where appropriate. Verification proceeds as in standard
Hoare logic, including formulating an appropriate invariant for
each loop. The most interesting invariant is for the main loop, shown in
Figure~\ref{fig:select-loop}; among other things, the invariant states that
\inlinec{head} points to a linked list \inlinec{l} of connection structures,
\inlinec{last_msg_store} points to a buffer storing a message \inlinec{m},
and the interaction tree under \inlinecoq{ITree} is an infinite loop of
\inlinecoq{select_loop_body} (Figure~\ref{fig:select-loop-model})) started on \inlinec{(l, m)}; the server address and
buffer size are constants.

Note that it is not immediate that the C loop body refines
\inlinecoq{select_loop_body}.  The former iterates over all ready
connections in \inlinec{process_connections}, while the latter works on only
one connection per iteration.  However, each iteration in
\inlinec{process_connections} is itself an iteration of
\inlinecoq{select_loop_body}, so the inner invariant carries the same
interaction tree. Conceptually, one iteration of the main loop in C
corresponds to multiple iterations of the model.

\section{Testing}\label{sec:testing}

Our overall approach to verifying software includes testing for errors in code and specifications before we invest too much effort in verification. For the swap server, we used QuickChick~\cite{quickchick}, a property-based testing tool in
Coq, to test both whether the C implementation satisfies the linear specification, and whether the
implementation model refines the linear specification. These tests help
establish confidence in all three artifacts.

\paragraph*{Test setup}
Our testbed consists of a simple hand-written client, the server to be
tested, and the linear specification that the server should satisfy. The
client opens multiple TCP connections to simulate multiple clients
communicating with the server over the network.

The testing process is straightforward: First, the client generates a
random sequence of messages along randomly chosen TCP connections.
The client then collects a trace of its interactions with the server---the
messages that it sent and the responses that it received in return on each
connection.
Finally, the checker attempts to ``explain'' this trace by enumerating all
the possible reorderings of this trace and
checking whether any of them is, in fact, a trace of the linear
specification. If such a trace is found, this test case passes, and another trace
is generated. If none of
the reorderings satisfies the specification, the tester reports that it
has found a counterexample.
Before actually displaying the counterexample, the tester
attempts to {\em shrink} it using a greedy search process modeled on the one
used in Haskell's QuickCheck tool, successively throwing away bits of the
counterexample and rechecking to see whether the remainder still fails.

We can also test that the implementation model refines the linear
specification.  The setup here is similar to the one for the C program, but
simpler because we can execute both the client and server
within a single Coq program rather than extracting a client from Coq and
running it with the server and a network.

\paragraph*{Testing the tester}
Although we did not find any bugs, we assessed the effectiveness of testing
using QuickChick's \emph{mutation testing} mode~\cite{mutation} to inject 12
different ``plausible bugs'' (of the sort commonly found in C: pointer errors,
bad initialization, off-by-one errors, \textit{etc.}) into the code and check
that each could be detected during testing.  The bugs are added to the C program
as comments marking a section of ``good code'' and a ``mutant'' that can be
substituted for it.  QuickChick performs this substitution for each of the
mutants in turn, generates random tests as usual, and reports how many tests it
took to find a counterexample for each of the mutants.

We analyzed the running time and number of tests needed to capture the bugs, by
repeating QuickChick for 29 times on each mutant.
For five of the 12 mutants,
the wrong behavior was caught by the very first test in each run. Six of the
mutants passed the first test in some runs, but always failed by the second
test.
The most interesting mutant was changing the return value of the
\inlinec{recv} call. 3/4 of the runs caught the bug within four rounds, but others
took up to nine rounds. This mutant sometimes causes the server not to respond,
which is trivially correct because our specification does not deal with
liveness. As a result, the tester discarded up to three thousand test cases where
the server did not respond, and ran for up to five minutes before failing.  The
other mutants could fail within 0.4 second with 95\% confidence.

It is hard to draw definite conclusions about the effectiveness of testing from
a case study of this size, but the fact that we are able to detect a dozen
different bugs, most quite quickly, is an encouraging sign that this approach to
testing will provide significant value as the codebase and its specification
become more complex.  Reports in the literature of property-based random testing
of similar kinds of systems (\textit{e.g.}, Dropbox~\cite{testing-dropbox}) are
also encouraging.

\section{Connecting to CertiKOS}
\label{sec:certikos}

A key pillar of the proof of correctness of the C implementation is the
specification of the socket operations such as \inlinec{send} and
\inlinec{recv}. We took these specifications as axioms when proving the
implementation model, but because we are running the server on top of
CertiKOS, which has its own formal specification, we should be able to go one
step better: we would like to prove that the socket operations as specified by CertiKOS
satisfy the axioms used in the VST proof.  This part of the case study is
still in progress; we report here on what we've achieved so
far and identify the challenges that remain.

\paragraph*{The Socket API in CertiKOS}
\label{sec:certikos-socket-api}

CertiKOS provides its own axiomatized specifications for the POSIX socket API.
Unlike VST specifications, which are expressed as Hoare triples, CertiKOS
specifications are written as state transition functions on the OS abstract state.
This state is a record with a field for each piece of real or ghost state that
the OS maintains. This includes, for example, buffers for received network messages,
or socket statuses. To provide a common language with VST for expressing allowable network
communications, we have modified CertiKOS' state to also include an ITree for each user process.

A function like \inlinec{recv} presents a challenge in that it depends
on nondeterministic behavior by the network, but the specification must be a deterministic
function. The standard solution used in CertiKOS is to parametrize the specification
by an ``environment context''~\cite{ccal}, which acts as a deterministic oracle that takes
a log of events and returns the next step taken by the environment.  Because the only
restriction on the environment context is that it is ``valid'' (\textit{e.g.}, for networks this
could mean that receive events always have a corresponding earlier send event),
properties proved about the specifications hold regardless of the particular choice of oracle.
Equipped with such a network oracle, the specification of \inlinec{recv} is fairly straightforward (Figure~\ref{fig:certikos-recv}).

\begin{figure}[t]
\begin{lstlisting}[style=customcoq, basicstyle=\scriptsize\ttfamily]
Definition recv_spec (fd maxlen : Z) (d : OSData)
  : option (OSData * Z) :=
  let pid := d.(curid) in
  (* Check that the ITree allows this behavior *)
  match ZMap.get pid d.(itrees) with
  | Vis (recv fd' maxlen') k =>
    if (fd = fd' && maxlen = maxlen') then
      (* Query the oracle for the next network message *)
      match net_oracle (ZMap.get pid d.(net)) with
      | RECV msg =>
        (* Take up to maxlen bytes *)
        let msg' := prefix maxlen msg in
        let len := length msg' =>
        (* Update the ITree based on len *)
        let res := if (len > 0) then inr (Some msg')
                   else if (len = 0) then inr None
                   else inl tt in
        let itree' := match res with
          | inl tt => ZMap.get pid d.(itrees)
          | inr msg => k msg end in
        (* Update the OS state and return len *)
        Some (d {itrees: ZMap.set pid itree' d.(itrees)}
                {rbuf: ZMap.set pid msg' d.(rbuf)}
                {net: RECV msg :: d.(net)}, len)
      | _ => None end
    else None
  | _ => None end.
\end{lstlisting}
\ifplentyofspace
\begin{lstlisting}[style=customcoq, basicstyle=\scriptsize\ttfamily]
  Definition sys_recv_spec (d: OSData) : option OSData :=
    (* Get the arguments from registers *)
    fd <- uctx_arg2_spec d ;;
    buf_vaddr <- uctx_arg3_spec d ;;
    len <- uctx_arg4_spec d ;;
    (d1, recv_len) <- recv_spec fd len d ;;
    (* Copy the contents of the kernel buffer to the
       user address *)
    d2 <- flatmem_copy_from_rbuf len buf_vaddr d1 ;;
    (* Set the return value *)
    d3 <- uctx_set_retval1_spec recv_len d2 ;;
    uctx_set_errno_spec E_SUCC d3.
\end{lstlisting}
\fi
\vspace*{-2.5ex}
\caption{CertiKOS specification of \inlinec{recv}.}
\label{fig:certikos-recv}
\end{figure}

\paragraph*{Bridging VST and CertiKOS memories}\label{sec:vst-and-certikos}

The other major gap between VST and CertiKOS is their treatment of memory. Both VST and CertiKOS build on CompCert's memory model to describe the state of memory, but the changes they make to it are unrelated and incompatible. VST builds a step-indexed model on top of CompCert memories~\cite{vst}, to
    allow for ``predicates in the heap''-based features, including
    recursive predicates and lock invariants. Hoare triples are interpreted as assertions on these step-indexed memories. On the other hand, the CompCert model corresponds to virtual memory, and treats independent memory allocations as belonging to separate, nonoverlapping ``blocks'', while CertiKOS uses a ``flat'' memory model in which there is only one block to more accurately represent the kernel's view of physical memory. To bridge this gap, we need to translate VST pre- and postconditions into assertions on ordinary, step-index-free CompCert memories (and vice versa), and transform predicates on multiple-block CompCert memories into predicates on CertiKOS's flat memories (and vice versa).

Performing this translation in general is an interesting research problem, but for this application, the specifications to be connected have a very particular form. The pre- and postconditions \inlinec{send} and \inlinec{recv} functions are each divided into two parts: a memory assertion on a single buffer, an array of bytes meant to hold the message, and an ITree assertion describing the external network behavior. This simplifies the task of connecting the VST and CertiKOS specs: we just need to relate the interaction tree to some component of the OS state, and translate an assertion on a single piece of memory into the flat memory model and back. (The other socket operations do not involve any changes to user memory, though they do modify kernel memory, which is abstracted to the C program via the \inlinecoq{SOCKAPI} predicate.)

We have explored this approach by sketching the correspondence between the VST specification of
\inlinec{recv} and its CertiKOS specification. We translated the VST pre-
and postcondition for \inlinec{recv} into step-index-free predicates on
CompCert memories and interaction trees by hand, and proved the correctness
of the translation using the underlying logic of VST. We then wrote
functions that transfer a single block of memory between the CompCert model
and the flat model, and adapted the CertiKOS OS component representing the
network state to use interaction trees, so that the two systems have a
common language to describe network operations. The network component of the
CertiKOS OS state is now a map that, for each user process, holds an
interaction tree describing the network communication that that process is
allowed to perform. Finally, we are in the process of proving that the CertiKOS
specification for \inlinec{recv} satisfies the step-index-free, flattened
versions of the VST pre- and postcondition. This gives us a path to validating
the axiomatized specifications of the socket API that we rely on for the
correctness of the C implementation: they can be substantiated by connection to
the (axiomatized) behavior of the socket operations in the underlying operating
system.

\section{Related Work}\label{sec:related-work}

\paragraph*{Interaction trees}

As mentioned in Section~\ref{sec:itrees}, our ``interaction trees'' are a
Coq-compatible variation of ideas found elsewhere. \citet{freer} present a similar
concept under the name ``freer monad''. It is proposed as an improvement over a
``free monad'' type, which one might hope to define in Coq as follows:

\begin{lstlisting}[style=customcoq, basicstyle=\scriptsize\ttfamily]
Inductive free (E : Type -> Type) (R : Type) :=
| Ret : R -> free E R
| Vis : E (free E R) -> free E R.  (* NOT PERMITTED!! *)
\end{lstlisting}

\noindent Unfortunately, the recursive occurrence of \inlinecoq{free} in the \inlinecoq{Vis}
constructor is not strictly positive, so this definition will be rejected by Coq. Thus
in a total language, the choice for the \inlinecoq{Vis} constructor to separate
the effect \inlinecoq{E X} from the continuation \inlinecoq{X -> itree E R} is
largely driven by necessity, whereas the work on freer monads proposes it as a matter of convenience and performance.

The \citet{mcbride-free} variant, which builds on earlier work by \citet{hancock-thesis}, is called the ``general monad.'' It is defined
inductively, and its effects interface replaces our single \inlinecoq{E : Type
  -> Type} parameter with \inlinecoq{S : Type} and a type family \inlinecoq{S ->
  Type} to calculate the result type.  It was introduced as a way to implement
general recursive programs in a total language (Agda), by representing recursive
calls as effects (\textit{i.e.}, \inlinecoq{Vis} nodes). Our coinductively defined interaction trees also
support a general (monadic) fixpoint combinator.

\citet{freespec} present the ``program monad'' to model components of
complex computing systems. Like the general monad, it is defined
inductively. Whereas our interpretation of ITrees is based on traces,
they use a coinductively defined notion of ``operational semantics'' to
provide the context in which to interpret programs, describing the state
transitions and results associated with method calls/effects.

Our choice to use coinduction and the \inlinecoq{Tau} constructor gives us a
way to account for ``silent''
(internal) computation steps, and hence allows us to semantically distinguish
terminating from silently-diverging computations (which is not easy with
trace-based semantics, at least not without adding a ``diverges'' terminal
component to some of the traces). Although liveness is explicitly not part of
our correctness specification in this project (the spec is conditioned
on there being visible output), it is conceivable to strengthen the
specifications and account for \inlinecoq{Tau} transitions as part of the C
semantics, which might allow one to prove liveness properties (although VST does
not currently support that).  However, there are also costs to working with
coinduction: our top-level programs are defined by \inlinecoq{CoFixpoint}, and
coinduction is generally not as easy to use in Coq as it could
be~\cite{coinduction, paco}.

\paragraph*{Verifying effectful systems}
A common approach to reasoning about effectful programs is to provide a model of
the state of the outside world, with access mediated strictly
through external functions. These functions may be given (possibly
non-deterministic) semantics directly~\cite{chlipala-network}, or indirectly
through an oracle~\cite{cal,feree-program}.  For example, in
\citet{feree-program}, external functions are called through a Foreign Function
Interface (FFI), and specification/verification is done with respect to an
instantiated FFI oracle that records external calls and defines the state of the
environment and the semantics of external functions. In their work,
a \texttt{TextIO} library was verified with respect to a model of the
file system.  Similarly, our specifications in terms of Hoare triples assume a
model of external socket API memory, \textit{i.e.}, the state under the
\inlinecoq{SOCKAPI} predicate, and describe how this state is transformed.

Stronger specifications of effectful programs can involve dynamics
(\emph{``what has happened''}) rather than statics
(\emph{``what is the final state''}).
In such cases, a model of the external state is commonly extended with
(or taken to be) a \emph{trace} or \emph{history} of past events, and
specifications involve these traces.
\citet{compcert, malecha-trace, cspec, ironfleet},~\textit{etc.} use this
approach.

Our specifications are based on interaction trees (which can be construed as
sets of traces), with one major difference: interaction trees specify
\emph{``what is allowed to happen''}. Rather than reasoning about \emph{lists}
of events that have occurred in the \emph{past}, our reasoning is based on the
\emph{trees} of events that are allowed to be produced in the \emph{future}. One
main advantage of using interaction trees is that it gives us a unifying
structure for specification, testing, and verification, as detailed in
\Cref{sec:itrees}. A similar underlying structure to interaction trees is used
as specifications of distributed systems in an early version of
F*~\citep{fstar}, but that work did not show how to use the structure for
testing or how to do refinement. \citet{ccal} use environment contexts to
specify past events as well as future events,
but rather than starting with all possible traces and consuming them, valid traces
are generated one event at a time by consulting an oracle.
Although using this step-based approach instead of explicitly coinductive ITrees leads
to different specification styles, it is possible to connect them as we discussed in
\Cref{sec:certikos}.

\paragraph*{Linearizability}

Network refinement is closely related to
linearizability~\cite{linearizability}, a correctness criterion for concurrent data
structures. A data structure implementation is \emph{linearizable} if,
for every possible collection of client threads, the behavior of the
data structure is indistinguishable from the behavior of a sequential
implementation of the
structure. \citet{Filipovic-linearizability} related
linearizability to contextual refinement. Network refinement is essentially this same idea of contextual refinement, but
with network effects playing the role of relaxed memory. Our network model closely resembles TSO, and network refinement is similar to TSO-linearizability~\cite{linearizability-tso}.

\paragraph*{Verifying networked servers}
In one early attempt at server verification, \citet{asv} verified security properties of the \texttt{thttpd} web
server, based on axiomatized C semantics. That work did not establish the functional
correctness of the web server, the axiomatic semantics was not testable, and it
did not consider the effects of network reordering.

IronFleet~\cite{ironfleet} is a methodology for verifying distributed
system implementations and it is similar to our approach in several ways: both verify the functional
correctness of a networked system; both use a ``one client at a time''
style specification at the top-level; and both verify the
correctness of a system implementation which interleaves its
operations via linearizability.
However, there are several major differences between IronFleet and our
work: (1) We are concerned with testing, as it allows us
to find implementation bugs early, and it also allows
us to use the same specification for blackbox-testing of existing
implementations. For these reasons, we choose the executable
interaction trees to represent the specification. IronFleet focuses instead
on reducing the burden of verification. It uses
non-executable state machines, and it relies on tool support such as near-real-time IDE-integrated feedback for rapid verification.
(2) Our work verifies C
implementations. VST and CompCert ensure that the properties we have
proved at the source-code level are preserved after the program has
been compiled to assembly code. IronFleet verifies programs written in
Dafny~\cite{dafny}, and extracts them to C\#. This means that both the
extraction engine and the .NET compiler must be trusted. The authors
of IronFleet also suggest an alternative strategy to reduce the
trusted computing base, by first translating the programs to assembly
code, and verifying the assembly code using an automatically
translated specification~\cite{ironclad}. However, that still requires
the specification translator to be trusted. (3) IronFleet is based on
UDP, while our works is based on TCP. Nevertheless, we both need to
consider packet reordering. The difference is that messages
will not be reordered on each individual connection. (4) IronFleet
uses TLA+~\cite{tla-plus} to prove liveness properties. The partial-correctness approach of separation logic makes it more difficult to reason about liveness.

CSPEC~\cite{cspec} is a framework for verifying concurrent
software. CSPEC focuses on reducing the number of interleavings a
verifier must consider. To do that, it provides a general verification
framework built on \emph{mover types}~\cite{lipton-reduction}. We
may be able to use mover types to simplify the process of proving network refinement.

Verdi~\cite{verdi} is a framework for verified distributed
systems that work under different fault and network models.  Verified System
Transformers transform a distributed system verified under one model
to one that works in another.  In particular, the Raft system
transformer~\cite{verdi-raft} transforms a given state machine (server) into a
distributed system of servers that synchronize state using Raft messages, over a
network that may drop, reorder, or duplicate messages. Any trace of Raft I/O
messages produced by the
distributed system can then be linearized to an I/O trace of the input state
machine. Distributed systems and transformers are written
in Coq and extracted to OCaml.

\citet{ridge09} verified the functional
correctness and linearizability of a networked, persistent message queue written
in OCaml using the HOL4 theorem prover.  In contrast to Verdi and
\citeauthor{ridge09}'s work, our methodology focuses on testing and verifying
C implementations, dealing with the full complexity of low-level
programming including memory allocation and pointer aliasing.

For simplicity, our work builds on a small subset of axiomatized TCP
specifications. A rigorous and experimentally-validated specification of TCP
can be found in \citet{TCP:tr, TCP:spec} and \citet{TCP:sl-spec}.

\paragraph*{Testing}

There is more research on testing linearizability of concurrent or distributed
systems than we can summarize here, including \citet{model-checking-linearizability, line-up,
  fuzzing-linearizability, testing-atomicity}.
Our work is distinguished by its focus on uniting testing and verification in
the same framework. QuickCheck's property-based testing methodology has been
shown to be useful in formal verification~\cite{quickchick,
  isabelle-quickcheck}. There are also many accounts of successfully applying
property-based random testing to real-world systems. For example,
\citet{quickcheck-database} used QuickCheck to test for race conditions in dets,
a vital component of the Mnesia distributed database system;
\citet{quickcheck-autosar} have applied the methodology to test the AUTOSAR
Basic Software for Volvo Cars, and \citet{testing-dropbox} have tested the
linearizability of Dropbox, the distributed synchronization service.

\section{Conclusions and Future Work}\label{conclusion}

Starting from a C implementation and a ``one client at a time'' specification of
swap server behavior, we have proved that every execution of the implementation
correctly follows the specification. The proof breaks down into layers of
refinements: from the C program to an implementation-level interaction tree, and
from there, via \textit{network refinement} to the linear interaction tree. We
use VST to verify the C code, pure Coq to relate the trees, QuickChick to test
our specifications and implementations, and CertiKOS to validate our
specifications of network communication. The result is a proof of the
correctness of the swap server from the linear specification down to the
interface between the C program and the operating system.

Although this work represents significant progress toward the Deep
Specification project's goal of formally-verified systems software, much remains
to be done.  The verification of the swap server has tested the limits of VST,
in terms of both scale and style of specifications. Previous VST verifications
were self-contained libraries, but this swap server interacts with the OS
through the socket API, requiring
us to develop new features (the external assertions) that should be useful for
verifying a variety of more realistic programs.
\ifplentyofspace The scale of this project forced us to debug and
streamline VST's existing automation.\fi

A clear next step is to fully verify the socket API used by the server, by
completing the proof that each VST socket axiom follows from the specification
of the corresponding operation in CertiKOS. Doing so will require several more
proofs along the lines of our verification of \inlinec{recv}, bridging the gap
between VST's step-indexed memory and CertiKOS's flat memory, as well as
defining a suitable C-level abstraction of the kernel memory related to the
socket operations. This will further extend the reach of our result, so that we rely only on the correctness of the operating system's model of the socket API.

Many real-world web servers are multi-threaded, handling requests from different
clients in separate threads. Some parts of our approach are already able to
handle concurrency: the top-level specification ITree should be sequential
regardless of the implementation, and VST and CertiKOS already support concurrent C
programs~\cite{mailbox,ccal}. Other parts will require adjustment: for instance, the
implementation model may need to explicitly represent the concurrency allowed in
the C program.

\subsection*{Acknowledgements} This work was funded by the National Science
Foundation's Expedition in Computing \emph{The Science of Deep Specification}
under the awards 1521602 (Appel), 1521539 (Weirich, Zdancewic, Pierce), and
1521523 (Shao), with additional support by the NSF projects \emph{Verified
  High Performance Data Structure Implementations}, award 1005849 (Beringer,
Mansky), and {\em Random Testing for Language Design}, award 1421243
(Pierce).  We are grateful to all the members of the DeepSpec project for
their collaboration and feedback, and we greatly appreciate the
reviewers' comments and suggestions.

\balance
\bibliography{ref}

\end{document}